\documentclass[twocolumn,preprintnumbers,amsmath,amssymb]{revtex4}

\usepackage{graphicx}
\usepackage{dcolumn}
\usepackage{bm}

\usepackage{amsmath}
\usepackage{amsfonts}
\usepackage{amssymb}

\newcommand{\beq}{\begin{equation}}
\newcommand{\ber}{\begin{eqnarray}}
\newcommand{\eeq}{\end{equation}}
\newcommand{\eer}{\end{eqnarray}}

\begin{document}

\preprint{SUNY BING 6/10/15}

\title{Diagrams and Parastatistical Factors for Cascade Emission of a Pair of Paraparticles}
\author{Charles A. Nelson}
\email{cnelson@binghamton.edu}

\author{Margarita Kraynova}
\author{Calvin S. Mera}
\author{Alanna M. Shapiro}
  
\affiliation{%
Department of Physics, State University of New York at Binghamton, Binghamton, New York 13902-6016\\
}%

\date{ June 10, 2015 } 

\begin{abstract}
The empirical absence to date of particles obeying parastatistics in high energy collider experiments might be due to their large masses, weak scale couplings, and lack of gauge couplings.  Paraparticles of order $p=2$ must be pair produced,  so the lightest such particles are absolutely stable and so are excellent candidates to be associated with dark matter and/or dark energy.   If there is a portal to such particles, from a new scalar $A_1$ boson they might be cascade emitted as a pair of para-Majorana neutrinos as in $A_1 \rightarrow A_2 \breve{\nu}_{\alpha}  \breve{\nu}_{\beta}$ or as a pair of neutral spin-zero paraparticles such as in $A_1 \rightarrow A_2 \breve{A}  \breve{B}$, where $\breve{B}$ is the anti-paraparticle to $\breve{A}$.  In this paper, for an assumed supersymmetric-like ``statistics portal" Lagrangian, the associated connected tree diagrams and their parastatistical factors are obtained for the case of order $p=2$ parastatistics.  These factors are compared with the corresponding statistical factors for the analogous emission of a non-degenerate or a 2-fold degenerate pair which obey normal statistics.  This shows that diagrams, and diagrammatic thinking, can be used in perturbatively analyzing paraparticle processes.  The parastatistical factor associated with each diagram does require explicit calculation.

\end{abstract}

\maketitle

\section{Introduction}

In the standard model all particles are either fermions or bosons which correspond to order $p=1$ parastatistics.   Identical fermions (bosons) occur only in the 1-dimensional totally antisymmetric column (totally symmetric row) representations of the permutation group.   Parastatistics  is a natural and simple generalization which includes the additional higher dimensional representations of the permutation group.  Fields and quanta obeying parastatistics are allowed in local relativistic quantum field theory [1-8].   Occasionally in this paper there are brief  summaries, such as in the appendices, so as not to assume that the reader has a quantum field theory background in parastatistics.

In this paper, we concentrate on order $p=2$ parastatistics, which is the simplest such generalization of normal Fermi and Bose statistics.   A simple consequence of order $p$  parastatistics is that up to $p$  identical parafermions (parabosons) can occupy a totally symmetric (antisymmetric) state, unlike for normal statistics.  More generally, identical parafermions (parabosons) of order $p$  occur in Young diagrams with at most $p$ columns (rows).  

Due to $p=2$ parastatistics, an even number of paraparticles must occur in the ``total external state" for a physical process,  so paraparticles must be pair produced and the lightest paraparticles are absolutely stable.  The ``total external state" consists of the particles in the initial state plus the final state.  Because of this absolute stability,  paraparticles of order $p=2$ are excellent candidates to be associated with dark matter and/or dark energy (accelerated expansion), given what is currently known from astrophysics and accelerator experiments.   

 If there is a ``statistics portal" from normal bosons and fermions to $p=2$ paraparticles at a high energy collider, then these particles might be emitted in a cascade process from a new scalar $A_1$ boson as a pair of para-Majorana neutrinos as in $A_1 \rightarrow A_2 \breve{\nu}_{\alpha}  \breve{\nu}_{\beta}$ or as a pair of spin-zero paraparticles such as in $A_1 \rightarrow A_2 \breve{A}  \breve{B}$, where $\breve{B}$ is the anti-paraparticle to $\breve{A}$.  
The paraparticles/parafields are denoted by a ``breve" accent.  All the new particles considered in this paper are assumed to be electromagnetically neutral with $\sim 100$ GeV to $\sim 2$ TeV scale masses.  The diagrammatic parastatistical factors are calculated for these two pair emission cascades because of their massive and unstable final $ A_{2} $ normal spin-zero boson,  versus the empirical difficulties for investigating a cascade to an almost massless final Majorana neutrino $\nu_2$ in $A_1 \rightarrow \nu_2 \breve{A}  \breve{\nu}$.   Depending on the unknown masses and coupling constants, these cascade processes might occur in the on-going experiments at the LHC with $\sqrt{s} \sim 13-14 $ TeV.  

As in the supersymmetric Wess-Zumino model [9], we assume that the portal Lagrangian densities for the cascade processes involve both a Majorana spin-$1/2$ field $\xi$ and a neutral complex spin-zero field $\mathcal A$ which respectively obey Fermi and Bose statistics, and parafermi and parabose counterparts  $ \breve{\xi}$ and  $ \breve{\mathcal A}$ which obey order $p=2$ parastatistics.   We consider this complex $\mathcal A$ field in the particle-antiparticle basis with corresponding quanta $A$ and $B$.  Similarly, $\breve A$ and $\breve B$ are the quanta for the complex spin-zero parabose $ \breve{\mathcal A}$ field. We will assume that there are two new $ A_{1,2} $ (with antiparticle $ B_{1,2}  $) bosons with $ m_1 > m_2>>0 $, that all mass values are at the $\sim 100$ GeV to  $\sim 2$ TeV scale, and that each of the cascade processes is  kinematically allowed.  We also assume that if not for their weak-scale portal associated couplings, the paraparticles would only interact gravitationally. Obviously, the 7 cascade processes considered in this paper are kinematically analogous to $\tau^{-} \rightarrow {\mu}^{-} \nu_{\mu} \nu_{\tau}$.  However, the $A_{1,2}$'s and $B_{1,2}$'s are spin-zero, so there do not exist useful polarization observables due to the cascading particle's spin, but concurrently there are fewer unknown possible covariant couplings.
 
 Using these Lagrangian densities, we perturbatively calculate the S-matrix elements for $A_1 \rightarrow A_2 \breve{\nu}_{\alpha}  \breve{\nu}_{\beta} , \cdots$ and $A_1 \rightarrow A_2 \breve{A}  \breve{B},  \cdots$.  We find that the tree diagrams for the associated connected amplitudes for cascade emission of a pair of paraparticles correspond to the same diagrams as in the case of the emitted pair obeying ordinary statistics, see Fig. 1 and others below.   While the diagrams are the usual covariant perturbative  ones, with the initial state on the left and the final state on the right, in labeling the virtual lines by $A$ or $B$, the displayed time-ordering has been assumed.  The arrows on the particle $A$ (antiparticle $B$) scalar boson lines are correspondingly forward (backward) in time. Since $\mathcal A$ is a complex field,  upon a time reversal of a time-ordered virtual line, exchange $A $ and $B$ label.  
 Unlike the spin-zero Higgs boson which is its own antiparticle, the neutral $\mathcal A$ field has distinguishable particle-antiparticle quanta. This same time-ordering property holds for time-ordered  $\breve{A}$ and $\breve{B}$  virtual lines associated with the $ \breve{\mathcal A}$ field.  In the figures, vertices and lines associated with the paraparticles are drawn heavy or ``dark."   There are also ``dark dots" on the external paraparticle legs which enables omission in the figures of an awkward ``breve" accent on the Weyl spinors.    In the case of $p=2$ parastatistics,  the parastatistical factors $c_p$ for the diagrams displayed are evaluated.

These $c_p$ factors in the $p=2$ para case are then compared with the analogous statistical factors  $c_d$ calculated for the amplitudes in the case of the emitted neutral pair obeying ordinary statistics and in the case when there is a hidden 2-fold degeneracy, for instance $A_1 \rightarrow A_2 {\nu}_{a,\alpha}  {\nu}_{a,\beta}$, where there are two kinds of emitted pairs ${\nu}_{a, \alpha}  {\nu}_{a, \beta}$ with $a=1,2$ the degeneracy index.  In the 2-fold degenerate case, as for a final particle polarization summation, this index $a$ is summed over to obtain the partial decay width.  The assumed portal Lagrangian densities considered  for these two comparison cases are analogous to those for the para case.

In agreement with what might have been anticipated by some readers, our explicit calculations show that for each diagram the statistical factor $c_p$ for order $p=2$ parastatistics, and hence the associated partial decay width, is the same as the $c_d$ statistical factor for such a 2-fold degeneracy.  
  
Section II  contains the supersymmetric-like Lagrangian densities assumed for these cascade processes.  It continues with the evaluations of the statistical factors $c_p$ in the para case and of the analogous factors $c_d$ in the cases of emission of a non-degenerate or a 2-fold degenerate pair obeying normal statistics.  Section III discusses the predictions for partial decay widths for these three cases.   Section IV has some concluding remarks.  

The relatively simple tri-linear relations for the creation and annihilation operators for an ``order $p=2$ family" of parafields are listed in  Appendix A.   

 \begin{figure}
\includegraphics[ trim= 420 0 0 250, scale=0.55 ]{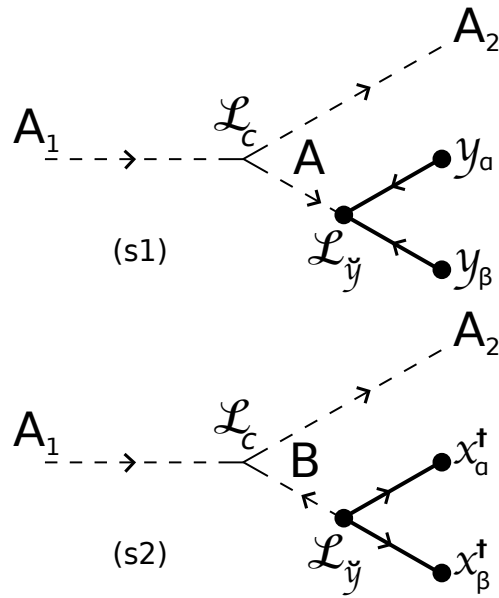}
\caption{\label{fig:epsart} Reading left-to-right, the first 2 of 4 diagrams for a cascade from a new scalar $A_1$ boson by emission of a pair of para-Majorana neutrinos $A_1 \rightarrow A_2 \breve{\nu}_{\alpha}  \breve{\nu}_{\beta}$. The virtual scalar $B$ is the antiparticle to $A$. Scalar bosons are denoted with thin dashed lines.  In the diagrams in this paper, ``dark dots" denote the portal Lagrangian vertices and the external $p=2$ paraparticles which have weak-scale portal associated couplings.  The ``dark" solid lines for the para-Majorana neutrinos are arrowed per the forward left-handed $x^{\dag}$ and backward right-handed  $y$ Weyl spinor, final state wave functions  as in DHM [10]. The Greek subscripts label the momenta and helicities of the para-Majorana neutrinos in (17) and (18).}
\end{figure}

\section{Cascade Processes with Emission of a Pair of Paraparticles}

\subsection{\label{sec:level2}Lagrangian densities}

For each of the interaction Lagrangian densities there is an explicit normalization of its coupling constant:  For fields obeying normal statistics, a factor of $({1}/{n!}) $ occurs when that field occurs to the $n$th power.  For a para Lagrangian density, two parafields occur in their appropriate commutator/anticommutator ordering,  see after (9), and also with an additional factor of $({1}/{2})$.  

While these are the usual normalizations associated with the identity of the fields in normal statistics and in parastatistics, these definitions are arbitrary.
However, these definitions of coupling constants are fixed and are used to calculate the statistical factors ($c_p$ and $c_d$) for each diagram/amplitude.  Any overall minus sign, or phase, is absorbed into the amplitude so $c_p, c_d  \geq 0$.  From the values obtained for these factors,  the consequences of alternate normalizations can be easily considered.  
The overall sign of each of the interaction Lagrangian densities has been arbitrarily chosen as minus.
  
Among the usual $p=1$ fields, we consider interactions as in the supersymmetric Wess-Zumino model  [9], but with unrelated weak scale coupling constants, so only slightly more general than in the supersymmetric limit.   We use the excellent supersymmetric formalism/notation of Dreiner-Haber-Martin (DHM) [10] with additional  ``breve" accents to denote the paraparticles/parafields.  The fields have their usual covariant momentum-expansions and normalizations in terms of their associated creation and annihilation operators [11].  The interaction densities involving only $p=1$ fields are
\ber
{\mathcal L}_{\mathcal Y}= - \frac{f}{2}  (\mathcal{A} \xi \xi +{\mathcal A}^{\dag} \bar{\xi} \bar{\xi})
\eer
\ber
{\mathcal L}_{C}= - \frac{t}{2} {  ( \mathcal A} ({\mathcal A}^{\dag})^{2 } +{\mathcal A}^2 {\mathcal A}^{\dag} )
\eer
\ber
{\mathcal L}_{q}= - \frac{F}{4} ({\mathcal A}^{\dag})^{2 }  {\mathcal A}^2  
\eer

For the cascade processes, we consider the following ``statistics portal" couplings between these p=1 fields and the p=2 fields, with anticommutator curly braces and commutator square brackets:
\ber
{\mathcal L}_{\breve{\mathcal Y} }= -  \frac{\breve f}{4}([\breve{\xi}, \breve{\xi}] \mathcal A 
+ {\mathcal A}^{\dag} [\bar{\breve{\xi}} ,\bar{\breve{\xi}}] )
\eer
\ber
{\mathcal L}_{2 \breve{c}}= - \frac{\breve{t}}{4} (  \{ { \breve{\mathcal A}} , {\breve{\mathcal A}} \}  {\mathcal A}^{\dag}  
+ {\mathcal A}  \{ {\breve{\mathcal A}}^{\dag} , {\breve{\mathcal A}}^{\dag} \} )
\eer
\ber
{\mathcal L}_{3 \breve{c}}= - \frac{\breve{T}}{2} ( {\mathcal A} +  {\mathcal A}^{\dag} )
 \{ {\breve{\mathcal A}} , {\breve{\mathcal A}}^{\dag} \} 
\eer
\ber
{\mathcal L}_{2 \breve{q}}= - \frac{\breve{F}}{8} (  \{ { \breve{\mathcal A}} , {\breve{\mathcal A}} \} ( {\mathcal A}^{\dag} )^2 
+  {\mathcal A} ^2  \{ {\breve{\mathcal A}}^{\dag} , {\breve{\mathcal A}}^{\dag} \} )
\eer
\ber
{\mathcal L}_{3 \breve{q}}= - \frac{\breve{G}}{4} \{ { {\mathcal A}} , {\mathcal A}^{\dag} \}
\{ {\breve{\mathcal A}} , {\breve{\mathcal A}}^{\dag} \}
\eer
\ber
{\mathcal L}_{\breve{\mathcal A} }= -  \frac{\breve j}{2}( \xi \{ \breve{\xi} , \breve{ \mathcal A } \}
+ \{  {\breve {\mathcal A}}^{\dag} , \bar{\breve{\xi}} \}   \bar{\xi} )
\eer
In these Lagrangian densities, the standard rules of paraquantization dictate the commutator/anticommutator ordering of the parafields. By ``paralocality" [3, 12] for fields obeying order $p=2$ parastatistics, two parafermi fields occur in a commutator ordering, whereas two parabose fields, or a parabose and a parafermi field, occur in an anticommutator ordering.  
Paralocality is a generalization of locality for parafields, see Appendix B.

For comparison, we also consider the case of cascade decays by pair emission fields ${\mathcal A}_a$ (neutral complex spin-zero)  and $\xi_a$ (Majorana spin-$1/2$) 
obeying respectively Bose and Fermi statistics, for instance $A_1 \rightarrow A_2 {\nu}_{a,\alpha}  {\nu}_{a,\beta}$.   For the degenerate case, the Lagrangian densities are analogous to the above portal ones:
\ber
{\mathcal L}^d_{\mathcal Y}= - \frac{f_d}{2}  (\mathcal{A} \xi_a \xi_a +{\mathcal A}^{\dag} \bar{\xi}_a \bar{\xi}_a)
\eer
\ber
{\mathcal L}^d_{2c}= -  \frac{t_d}{2} ( {\mathcal A}^{\dag}  {\mathcal A}_a {\mathcal A}_a  
+ {\mathcal A}^{\dag}_a {\mathcal A}^{\dag}_a {\mathcal A} ) 
\eer
\ber
{\mathcal L}^d_{3c}= - {T_d}  ( {\mathcal A}  {\mathcal A}_a {\mathcal A}^{\dag}_a  
+{\mathcal A}_a {\mathcal A}^{\dag}_a {\mathcal A}^{\dag} ) 
\eer
\ber
{\mathcal L}^d_{2q}= -  \frac{F_d}{4} (   {\mathcal A}_a {\mathcal A}_a ( {\mathcal A}^{\dag} )^2
+  {\mathcal A}^2 {\mathcal A}^{\dag}_a {\mathcal A}^{\dag}_a )
\eer
\ber
{\mathcal L}^{d}_{3q}= - G_{d} ( {\mathcal A}_a {\mathcal A}^{\dag}_a )  ( {\mathcal A} {\mathcal A}^{\dag} )
\eer
\ber
{\mathcal L}^d_{\mathcal A}= - {j_d}  ( \xi \xi_a \mathcal{A}_a +{\mathcal A}^{\dag}_a \bar{\xi}_a \bar{\xi})
\eer
For the case of 2-fold degeneracy, the degeneracy index $(a=1,2)$ is summed over in these densities.

The interaction Lagrangian densities which do not occur in the cascade processes calculated in this paper are (1) and (3), and in the paraparticle portal case (9) and its analog (15)  in the degenerate case.   However, the empirically difficult to observe cascade to an almost massless final Majorana neutrino $\nu_2$ in $A_1 \rightarrow \nu_2 \breve{A}  \breve{\nu}$ does involve both (1) and the portal coupling (9), and its degenerate counterpart (15).  

\subsection{\label{sec:level2}Parastatistical factors for 7 cascade processes}

The above interaction Lagrangian densities have a particle-antiparticle transformation symmetry such that the results obtained for each cascade also hold for the cascade obtained by transforming all $ {A}_i \leftrightarrow   {B}_i $  and $ \breve{A} \leftrightarrow   \breve{B}$. For instance, the parastatistical factors are the same for $A_1 \rightarrow A_2 \breve{A}  \breve{B}$ and $B_1 \rightarrow B_2 \breve{B}  \breve{A}$.    For the normal statistics cascades involving ${\mathcal A}_a$  and $\xi_a$, there is the analogous transformation of all $ {A}_i \leftrightarrow   {B}_i $  and $A_{a} \leftrightarrow   B_{a}$.  Consequently, the statistical factors $c_p$ and $c_d$ obtained below for the diagrams in the scalar $A_1$ decay process are the same as for the associated antiparticle $B_1$ decay process because of this particle-antiparticle transformation symmetry.


\subsubsection{Emission of a pair of para-Majorana neutrinos: $A_1 \rightarrow A_2 \breve{\nu}_{\alpha}  \breve{\nu}_{\beta}$ and $A_1 \rightarrow B_2 \breve{\nu}_{\alpha}  \breve{\nu}_{\beta}$}

In this paper the evaluations of the S-matrix elements only involve processes with a pair of final paraparticles.  We calculate the associated amplitudes in the ``occupation number basis" for a specific ordering of the two particles in the pair and then by addition or subtraction, construct the corresponding amplitudes in the ``permutation group basis"  [2] to obtain the physical amplitudes for the pair of paraparticles.    In these evaluations, calculating in the occupation number basis halves the number of terms, versus using the permutation group basis, and a simple relabeling in the final expression gives the amplitude for the opposite ordering of the final two paraparticles.    This  distinction between fundamental bases in parastatistics  is explicitly and simply explained below in the context of the calculation of the amplitudes for $A_1 \rightarrow A_2 \breve{\nu}_{\alpha}  \breve{\nu}_{\beta}$ associated with the two Fig. 1 diagrams.  This leads to the discussion in the text of the two physical permutation group basis final states of (19) below.  

In canonical quantum field theory, for particles obeying normal statistics there is a successful normal ordering procedure for correctly ordered Lagrangian densities which is used in the perturbative evaluation of S-matrix elements [13].  This procedure discards various diagrams and yields results for the standard model which are currently in highly precise agreement with experimental data.  However, this procedure has not been generalized for paraparticles.  Nevertheless, as shown in this paper, knowing from purely $p=1$ quanta the canonical assembly of contributions from the perturbative evaluation into physical amplitudes, we find that it is straight-forward to proceed analogously by hand for $p=2$ paraparticles using the above paraquantized Lagrangian densities:   We require each field in ${\mathcal L}_{int}$ to contract with a field in a different ${\mathcal L}_{int}$ or with a particle in the initial or final states.  This omits disconnected diagrams and ones with a single ${\mathcal L}_{int}$ term self-contraction.  
In this context, it is important to note that there are highly non-trivial signs associated with this diagrammatic application of the tri-linear quantization relations and of the paralocal Lagrangian densities involving order $p=2$ fields.   Clearly, the two crucial tests of this systematic diagrammatic evaluation of paraparticle S-matrix elements will be whether it generalizes in perturbative quantum field theory and whether the resultant amplitudes do indeed agree with experiment.

(i) We first consider a cascade from a new $A_1$ boson by emission of a pair of para-Majorana neutrinos $A_1 \rightarrow A_2 \breve{\nu}_{\alpha}  \breve{\nu}_{\beta}$:  

From  
${\mathcal L}_{\breve{\mathcal Y} }$ and ${\mathcal L}_{C}$ there is the following time-ordered product
\ber
S_{fi}= (i  )^2  \int d^4 x_1  \int d^4 x_2 ~ 
\theta( t_1 - t_2) \nonumber \\
{}_A\!\!< A_l \breve{\nu}_{\alpha}  \breve{\nu}_{\beta}| \{ 
{\mathcal L}_{\breve{\mathcal Y} }(x_1)    {\mathcal L}_{C}(x_2)     +      {\mathcal L}_{C}(x_1) {\mathcal L}_{\breve{\mathcal Y} }(x_2)
\} |A_k>  \nonumber\\
\eer
The final state has the $ \breve{\nu}_{\alpha}  \breve{\nu}_{\beta} $ paraparticle operators ordered as  
$ | A_l  \breve{\nu}_{\alpha}  \breve{\nu}_{\beta} >_A = \frac{1}{2} l^{\dag} \alpha^{\dag} \beta^{\dag} |0> $ in the occupation number basis.   The role of the subscript on the ket-state (bra-state) is to denote the place-position order [14].  We are labeling one place-position order ``A" ($ \alpha^{\dag} \beta^{\dag}$) and its orthogonal counterpart ``B" ($ \beta^{\dag} \alpha^{\dag} $),  where the ordering of the operators is reversed  $ | A_l  \breve{\nu}_{\alpha}  \breve{\nu}_{\beta} >_B = \frac{1}{2} l^{\dag}  \beta^{\dag} \alpha^{\dag} |0> $.   In $p=2$ parastatistics there are the tri-linear relations instead of the usual bi-linear ones, so these A  and B  orderings in the occupation number basis must be distinguished.   

We label the $A_l$ creation operator by $l^{\dag}$.  Notice the essential and easy to forget two paraparticles' factor of $(\frac{1}{\sqrt{2}})^2$ in the state norm due to the vacuum condition 
$ {\alpha}_k {\alpha}_l^{\dag}                      | 0 > = 2 \delta_{kl}  | 0 > $.  These extra paraparticle normalization factors occur because our calculations depend on the ``arbitrary $p$ normalization," see Appendix A.

By writing the fields of the A-ordered final state in (16) in terms of their positive- and negative- frequency parts, and then using the $p=2$ tri-linear relations for the paraquanta, we obtain amplitudes corresponding to the two connected tree diagrams shown in Fig. 1.  See Appendix C for  $p=2$ normalization details.

From (16), the (s1) amplitude [11] for the A-ordered final state is
\ber
- i \mathcal{M}^{(s1)}_A = \{  1 \} (i t)( i \breve{f} ) \frac{i}{q^2_A -{m^2} + i \epsilon}  y ^ A ( \vec{p}_\alpha,\lambda_{\alpha} ) 
y_A  ( \vec{p}_\beta,\lambda_{\beta} )  \nonumber\\
\eer
with A being the 2-valued summed index for the commuting two-component ``right-handed Weyl" spinor, final state wave function $ y_A$ of DHM [10], and the (s2) amplitude is
\ber
- i \mathcal{M}^{(s2)}_A = \{  1 \} (i t)( i \breve{f} ) \frac{i}{q^2_B -{m^2} + i \epsilon}  x^{\dag}_{\dot{A}} ( \vec{p}_\alpha,\lambda_{\alpha} ) 
x^{\dag \dot{A}} ( \vec{p}_\beta,\lambda_{\beta} ) 
 \nonumber \\
\eer
with A-dot being the 2-valued summed index for the commuting, conjugate ``left-handed Weyl" spinor, final state wave function ${x^{\dag}}_{\dot A}$.  Because of our usage of Greek letters for parafermions, such as in the tri-linear relations in Appendix A, we use undotted and dotted capital Roman letters for these 2-valued Weyl indices in place of  the lowercase Greek letters in DHM.   

As in DHM, the arrows on the two-component spinor lines correspond to fields with undotted (dotted) indices flowing into (out of) any vertex.  The direction of an arrow, versus a vertex, for either a spin-zero or a spin-1/2 line is unchanged upon any time reordering of a displayed diagram.  From the Lagrangian densities (1), (4), (9) and (10), (15), the arrows on the lines for the scalar fields $\mathcal A$ and $ \breve{\mathcal A}$ are also into (out of) any vertex per the common into (out of) direction of the two spinor lines.

\begin{figure}
\includegraphics[ trim= 420 0 0 250, scale=0.55 ]{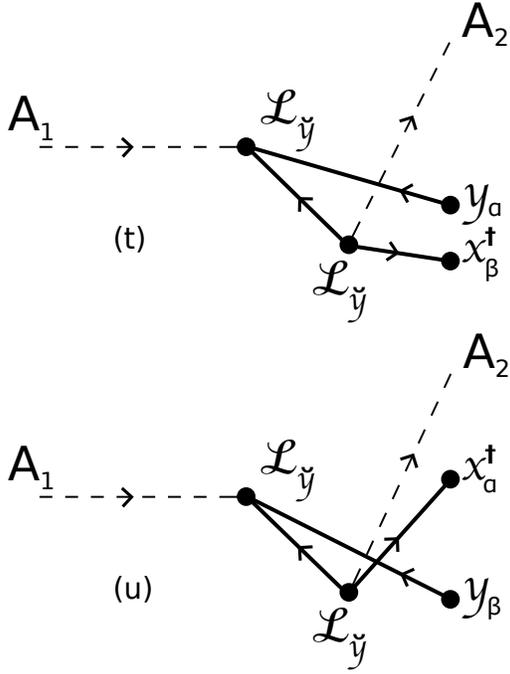}
\caption{\label{fig:epsart} The remaining 2 diagrams for $A_1 \rightarrow A_2 \breve{\nu}_{\alpha}  \breve{\nu}_{\beta}$.}
\end{figure}

To maintain simplicity of the expressions for the matrix elements,  we  omit the associated mixing matrices between the mass eigenstates and the interaction eigenstates for the external $   A_{1,2} $ bosons. In this paper, the amplitude/diagrammatic normalization is for a single $\mathcal A $, $ \breve{\xi}$, or $\breve{\mathcal A}$ in the virtual propagators.  Also, while the standard model lacks sufficient CP-violation for the observed baryon and lepton asymmetries of the universe, we omit explicit CP-violation formalism and possible mixing of the $A_{1,2}$ with the $B_{1,2}$ bosons.

For comparison, in the case with pair emission fields ${\mathcal A}_a$  and $\xi_a$  obeying the usual Bose and Fermi statistics, the same amplitudes for (s1) and (s2) are obtained
for the process $A_1 \rightarrow A_2 {\nu}_{a,\alpha}  {\nu}_{a,\beta}$ with $ | A_l  {\nu}_{a,\alpha}  {\nu}_{a,\beta} > =  l^{\dag} \alpha_a^{\dag} \beta_a^{\dag} |0> $ except in place of the parastatistical factor $\{  1 \} ( t \breve{f} )$, there is instead a factor of  $\{ 1 \} (t  f_d )$, where the respective statistical factors ${c_p} $ and ${c_d}$ are given in the curly braces.   In writing these statistical factors times coupling constants, we omit each $( i)$  associated with the $ i {\cal{L}}_{int}$ vertex. This is the comparison amplitude for all fields obeying ordinary statistics for the Lagrangian densities given in (10-15).   In the 2-fold degenerate case where there are two kinds of emitted pairs ${\nu}_{a,\alpha}  {\nu}_{a,\beta}$, calculation of the partial decay width requires a factor of 2 due to summing over the two final degenerate channels.

\begin{figure}
\includegraphics[ trim= 420 0 0 250, scale=0.55 ]{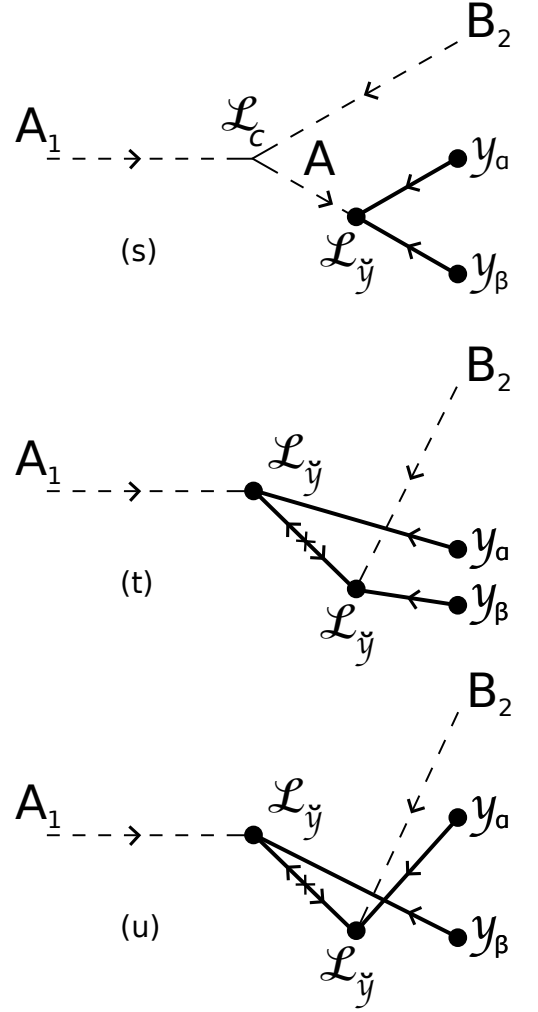}
\caption{\label{fig:epsart}  The 3 diagrams for $A_1$ cascading to $B_2$ by emission of a pair of para-Majorana neutrinos $A_1 \rightarrow B_2 \breve{\nu}_{\alpha}  \breve{\nu}_{\beta}$.    There is a para-Majorana mass insertion in the $(t)$ and $(u)$ diagrams. }
\end{figure}

For the orthogonal B-ordered final state, the same amplitude for (s1), and similarly for (s2), is obtained but with an opposite overall sign in comparison to the A-ordered final state, so that the permutation group basis amplitudes $\mathcal{M}^{(s1)} $ and $\mathcal{M}^{(s2)} $ for the symmetric/antisymmetric final states
\ber
 | A_l  \breve{\nu}_{\alpha}  \breve{\nu}_{\beta} >_{sym,asym} = \frac{1}{\sqrt{2}}  ( | A_l  \breve{\nu}_{\alpha}  \breve{\nu}_{\beta} >_A \pm | A_l  \breve{\nu}_{\alpha}  \breve{\nu}_{\beta} >_B ) \nonumber\\
 \eer
  are respectively zero and $\sqrt{2}$ times those for the A-ordering.    Hence, from the values of the statistical factors ${c_p} $ and ${c_d}$,  if these were the only two diagrams, upon summing over the two permutation basis final states for the decay process $A_1 \rightarrow A_2 \breve{\nu}_{\alpha}  \breve{\nu}_{\beta}$ the partial decay width would be twice that for the corresponding normal statistics process $A_1 \rightarrow A_2 {\nu}_{a,\alpha}  {\nu}_{a,\beta}$ with a non-degenerate pair.  However, the $p=2$ partial decay width would be the same as that for the case of emission of two kinds of pairs ${\nu}_{a,\alpha}  {\nu}_{a,\beta}$ due to summing over these two degenerate channels.

For the $A_1 \rightarrow A_2 \breve{\nu}_{\alpha}  \breve{\nu}_{\beta}$ cascade, there is also a contribution from   
$({\mathcal L}_{\breve{\mathcal Y} })^2$ which corresponds to the two diagrams in Fig. 2. 
Again, for each diagram, the B-ordering gives the same amplitude, but with opposite overall sign versus the A-ordering.  Also, again for  the A-ordering, the expressions associated with the diagrams are proportional in the case of paraparticles and the $p=1$ (normal statistics) case of non-degenerate Majorana neutrinos. The contribution of the $(u)$ diagram is minus that of the $(t)$ diagram with $\alpha$ and $  \beta$ exchanged.  In the para case, the $(t)$ diagram has a factor of $\{  1 \} ( \breve{f} )^2 $, and in the $p=1$ case there is a factor of $\{  {1} \} (f_d )^2$ instead.  In the evaluation for the para case, there is a factor of $2$ which arises from transforming the position space propagator vacuum expectation value to momentum space, see Appendix C.

(ii) As shown in Fig. 3, there is a similar cascade from $A_1$ to the antiparticle $B_2$ by the emission of a pair of para-Majorana neutrinos, $A_1 \rightarrow B_2 \breve{\nu}_{\alpha}  \breve{\nu}_{\beta}$:

\begin{figure}
\includegraphics[ trim= 420 0 0 250, scale=0.55 ]{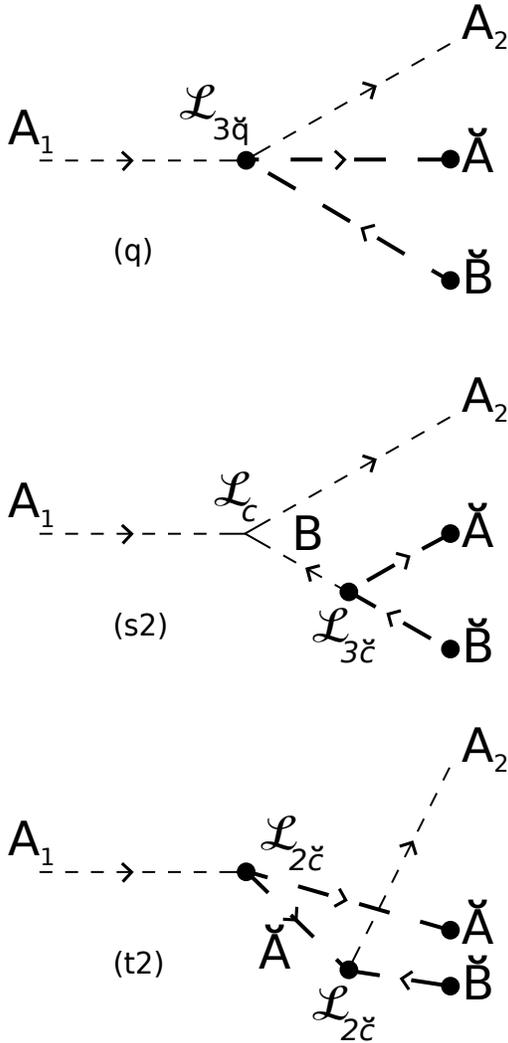}
\caption{\label{fig:epsart} First 3 of 6 diagrams for the cascade $A_1 \rightarrow A_2 \breve{A}  \breve{B}$ by emission of a particle-antiparticle pair of scalar paraparticles $\breve{A} \breve{B}$.   $\breve{B}$ is the anti-paraparticle to $\breve{A}$. These scalar paraparticles are denoted by ``dark" dashed lines with forward (backward) in time arrows for the particle $\breve{A}$ (antiparticle $ \breve{B}$).}
\end{figure}

For each diagram, for the A-ordering the para amplitude is proportional to that obtained in the case of ordinary fermion Majorana neutrinos.  Also for each diagram, the B-ordered expression is of opposite sign to that of the A-ordering, so the permutation group basis amplitude is again the asymmetric one.

From ${\mathcal L}_{\breve{\mathcal Y} }$ and ${\mathcal L}_{C}$, for the A-ordering there is a single $(s)$ diagram with a parastatistical factor of $\{  1 \} ( t \breve{f} )$.    For the analogous $p=1$ cascade $A_1 \rightarrow B_2 {\nu}_{a,\alpha}  {\nu}_{a,\beta}$, there is a factor of $\{  1 \} (t  f_d )$.  The contribution from  $({\mathcal L}_{\breve{\mathcal Y} })^2$ involves a para-Majorana mass insertion contribution.  The amplitude for the $(u)$ diagram is again minus that of the $(t)$ diagram with $\alpha $ and $ \beta$ exchanged.  For the $(t)$ diagram, in the para case there is a factor of $\{  1 \} ( \breve{f} )^2 $ and correspondingly in the $p=1$ fermion case a factor of $\{  {1} \} (f_d )^2$.

\subsubsection{Emission of a pair of scalar paraparticles:  
\newline $A_1 \rightarrow A_2 \breve{A}  \breve{B}$, $A_1 \rightarrow B_2 \breve{A}_3 \breve{A}_4$, $\cdots$}

In the remaining 5 cascade processes, $A_1 \rightarrow A_2 \breve{A}  \breve{B}$, $A_1 \rightarrow B_2 \breve{A}_3 \breve{A}_4$, $\cdots$, a pair of scalar paraparticles are emitted.  For each process, the obtained A-ordered amplitudes can again be considered in terms of its covariant diagrams which are displayed in the figures.  These A-amplitudes in the para case are again proportional to those in the non-degenerate case in which there is a scalar pair emitted.  In the following, for each diagram the respective statistical factors ${c_p} $ and ${c_d}$ are listed.

\begin{figure}
\includegraphics[ trim= 420 0 0 250, scale=0.55 ]{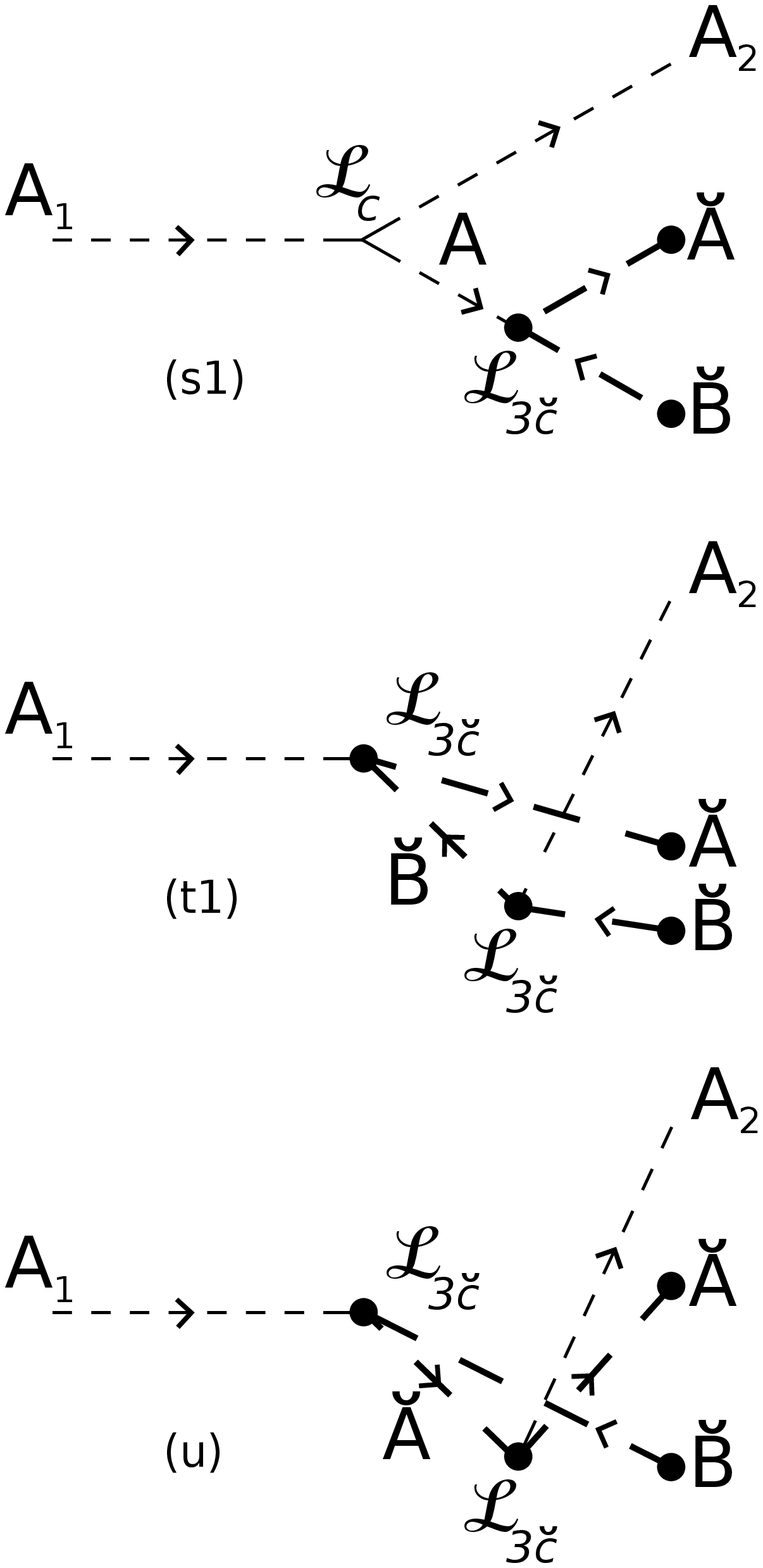}
\caption{\label{fig:epsart} The remaining 3 diagrams for $A_1 \rightarrow A_2 \breve{A}  \breve{B}$.}
\end{figure}

For each diagram the same amplitudes are obtained for the A-ordered and B-ordered final states.   Therefore, in the permutation group basis the associated symmetric final state has an amplitude of $\sqrt{2}$ times that for the A-ordering, and the amplitude for the antisymmetric final state vanishes.  For the first cascade $A_1 \rightarrow A_2 \breve{A}  \breve{B}$ with emission of a particle-antiparticle pair of paraparticles, the symmetric/antisymmetric final states are 
\ber
 | A_{2,l} \breve{A}  \breve{B}  >_{sym,asym} = \frac{1}{\sqrt{2}}  ( | A_{2,l}  \breve{A}  \breve{B} >_A \pm | A_{2,l}  \breve{A}  \breve{B}>_B ) \nonumber \\
 \eer
with A-ordering and B-ordering of the kets $ | A_{2,l} \breve{A}  \breve{B} >_A = \frac{1}{2} l^{\dag} A^{\dag} B^{\dag} |0> $ and $ | A_{2,l}  \breve{A}  \breve{B} >_B = \frac{1}{2} l^{\dag} B^{\dag} A^{\dag} |0> $.

(i) Fig. 4 shows the first 3 diagrams for the cascade $A_1 \rightarrow A_2 \breve{A}  \breve{B}$.

Fig. 5 shows the remaining 3 diagrams: 

\begin{figure}
\includegraphics[ trim= 420 0 0 250, scale=0.55 ]{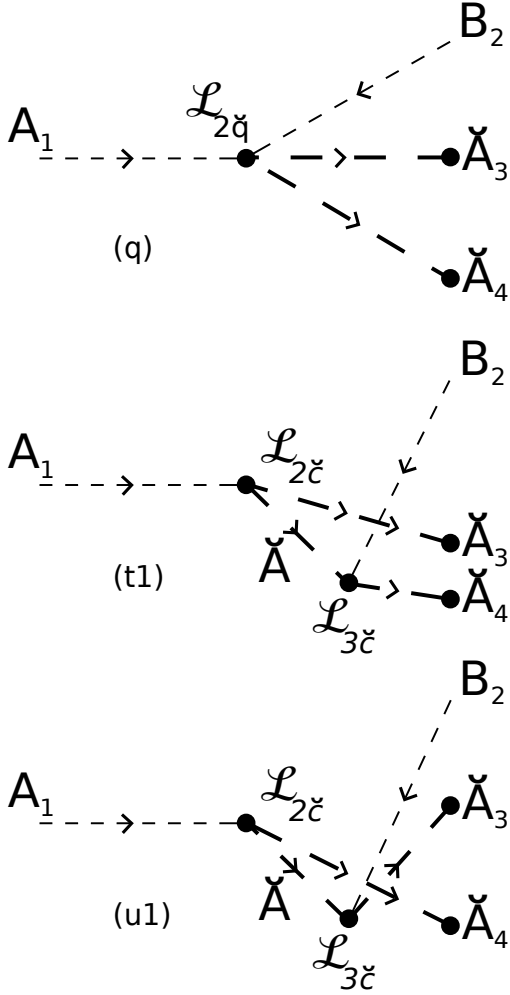}
\caption{\label{fig:epsart} First 3 of 6 diagrams for the cascade $A_1 \rightarrow B_2 \breve{A}_3 \breve{A}_4 $ by emission of a pair of scalar paraparticles $\breve{A}_3 \breve{A}_4$.  The subscripts on $\breve{A}_3$ and $ \breve{A}_4$ are for momentum labeling which distinguishes the two identical parabosons. }  
\end{figure}
\begin{figure}
\includegraphics[ trim= 420 0 0 250, scale=0.55 ]{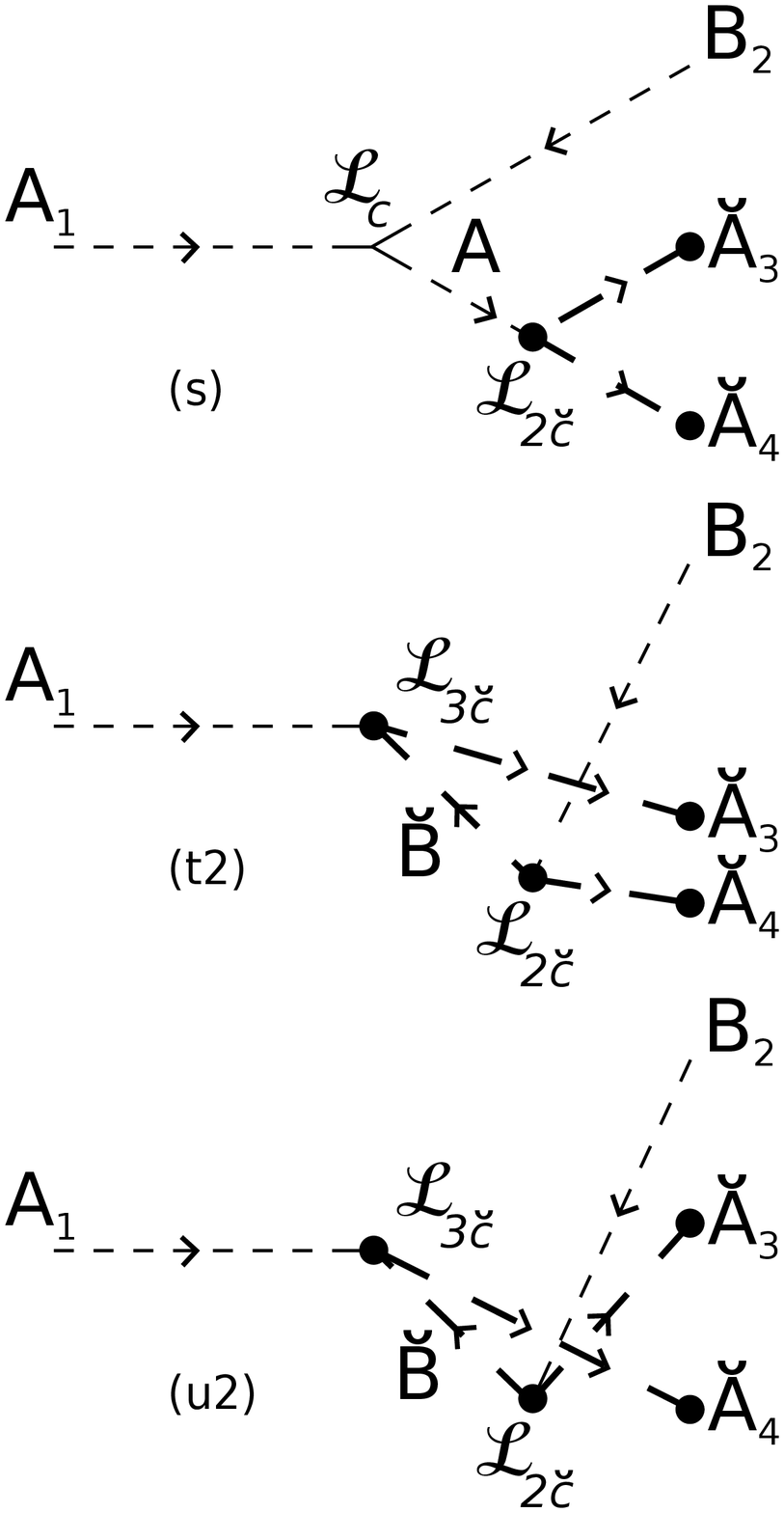}
\caption{\label{fig:epsart} The remaining 3 diagrams for $A_1 \rightarrow B_2 \breve{A}_3 \breve{A}_4 $.} 
\end{figure}

From ${\mathcal L}_{3 \breve{q}}$, there is  the $(q)$ diagram with a factor of
$\{  1 \} ( - \breve{G} )$ versus $\{  1 \} ( - G_d )$. The minuses occur here because we omit each $( i)$  associated with the $ i {\cal{L}}_{int}$ vertex.  From $ {\mathcal L}_{C} $ and ${\mathcal L}_{3 \breve{c}} $,  the $(s1)$ and $(s2)$ diagrams each have a factor of $\{1 \} ( t \breve{T} )$ versus $\{  1 \} ( t T_d )$.  From the $({\mathcal L}_{3 \breve{c}})^2 $ contribution,  the $(t1)$ and $(u)$ diagrams each have a factor of $\{ 1 \} ( \breve{T} )^2$ versus $\{  1 \} ( T_d )^2$.  
Interestingly, there is only a single diagram contribution from $({\mathcal L}_{2 \breve{c}} )^2$.  This $(t2)$ diagram has a factor of $\{  1 \} ( \breve{t} )^2$ versus $\{  1 \} ( t_d )^2$.

(ii) The analogous cascade from $A_1$ to the antiparticle $B_2$, $A_1 \rightarrow B_2 \breve{A}_3 \breve{A}_4 $, has the 6 diagrams shown in Figs. 6 and 7:  

From ${\mathcal L}_{2 \breve{q}}$, there is the $(q)$ diagram with a factor of
$\{  1 \} ( -  \breve{F} )$ in the para case versus a factor of $\{  1 \} ( -  F_d )$ in the boson case. From $ {\mathcal L}_{C} $ and ${\mathcal L}_{2 \breve{c}} $,  the $(s)$ diagram has a factor of  $\{ 1  \} ( t \breve{t} )$ versus $\{  1 \} ( t t_d )$.  As shown, the remaining four diagrams arise from $ {\mathcal L}_{2 \breve{c}}$ and ${\mathcal L}_{3 \breve{c}} $.  They are $(t1)$, $(u1)$, $(t2)$, and $(u2)$.  Each has a factor of  $\{  1 \} ( \breve{t} \breve{T} )   $ versus $ \{  1 \} ( t_d  T_d ) $.

(iii) The cascade from $A_1$ to $A_2$ by $A_1 \rightarrow A_2 \breve{A}_3 \breve{A}_4 $  has the diagrams shown in Fig. 8:  
\begin{figure}
\includegraphics[ trim= 420 0 0 250, scale=0.55 ]{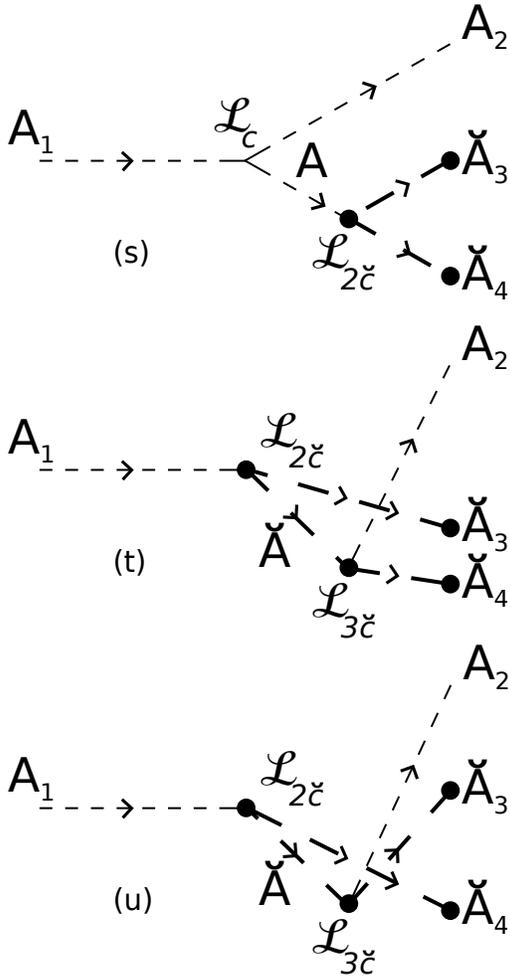}
\caption{\label{fig:epsart} 
The 3 diagrams for the cascade $A_1 \rightarrow A_2 \breve{A}_3 \breve{A}_4 $ by emission of a pair of scalar paraparticles $\breve{A}_3 \breve{A}_4$.} 
\end{figure}

From ${\mathcal L}_{C} $ and ${\mathcal L}_{2 \breve{c}} $,  the $(s)$ diagram has a factor of  $\{ 1 \} ( t \breve{t} )$ versus $\{  1 \} ( t t_d )$.  From $ {\mathcal L}_{2 \breve{c}} $ and ${\mathcal L}_{3 \breve{c}} $,  the $(t)$ and $(u)$ diagrams each has a factor of $\{ 1 \} ( \breve{t} \breve{T} )$ versus $\{  1 \} ( t_d T_d )$.

(iv) If instead there is emission of antiparticle pair $\breve{B}_3 \breve{B}_4$ via the cascade $A_1 \rightarrow A_2 \breve{B}_3 \breve{B}_4 $,
there are the diagrams shown in Fig. 9:  
\begin{figure}
\includegraphics[ trim= 420 0 0 250, scale=0.55 ]{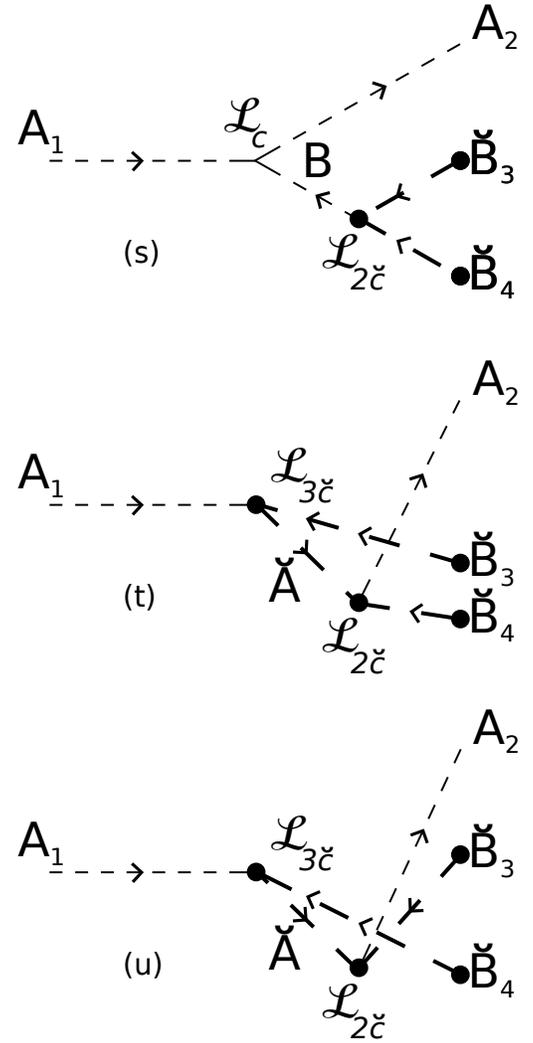}
\caption{\label{fig:epsart} The 3 diagrams for the cascade $A_1 \rightarrow A_2 \breve{B}_3 \breve{B}_4 $ by emission of an antiparticle pair of scalar paraparticles $\breve{B}_3 \breve{B}_4$.}
\end{figure}

From ${\mathcal L}_{C} $ and ${\mathcal L}_{2 \breve{c}} $,  the $(s)$ diagram has a factor of  $\{ 1 \} ( t \breve{t} )$ versus $\{  1 \} ( t t_d )$.  From  $ {\mathcal L}_{2 \breve{c}} $ and ${\mathcal L}_{3 \breve{c}} $,  the $(t)$ and $(u)$ diagrams each have a factor of $\{ 1  \} ( \breve{t} \breve{T} )$ versus $\{  1 \} ( t_d T_d )$.  The  interaction vertices $ {\mathcal L}_{2 \breve{c}} $ and ${\mathcal L}_{3 \breve{c}} $  in the $(t)$ and $(u)$ diagrams are exchanged in Fig. 9 for emission of $\breve{B}_3 \breve{B}_4$ versus those in Fig. 8 for emission of $\breve{A}_3 \breve{A}_4$. 

(v) For the cascade $A_1 \rightarrow B_2 \breve{A}_3 \breve{B}_4 $ by emission of $\breve{A}_3 \breve{B}_4$ there are the diagrams in Fig. 10:

From ${\mathcal L}_{C} $ and ${\mathcal L}_{3 \breve{c}} $,  the $(s)$ diagram has a factor of  $\{1 \} ( t \breve{T} )$ versus $\{  1 \} ( t T_d )$.  From  second order in $ {\mathcal L}_{3 \breve{c}} $,  the $(t)$ and $(u)$ diagrams each have a factor  of $\{ 1 \} (\breve{T} )^2$ versus $ \{  1 \} ( T_d )^2 $.  

 As briefly explained in Appendix D, the same parastatistical factor $c_p$ (as above) is obtained for each diagram in the alternate $p=2$ normalization of Green and Volkov [1] for the tri-linear commutation relations.  To achieve these same $c_p$ values, there is a necessary rescaling of each of the portal coupling constants in (4-9) by $ g_i \rightarrow 2 g_i $. 

\begin{figure}
\includegraphics[ trim= 420 0 0 250, scale=0.55 ]{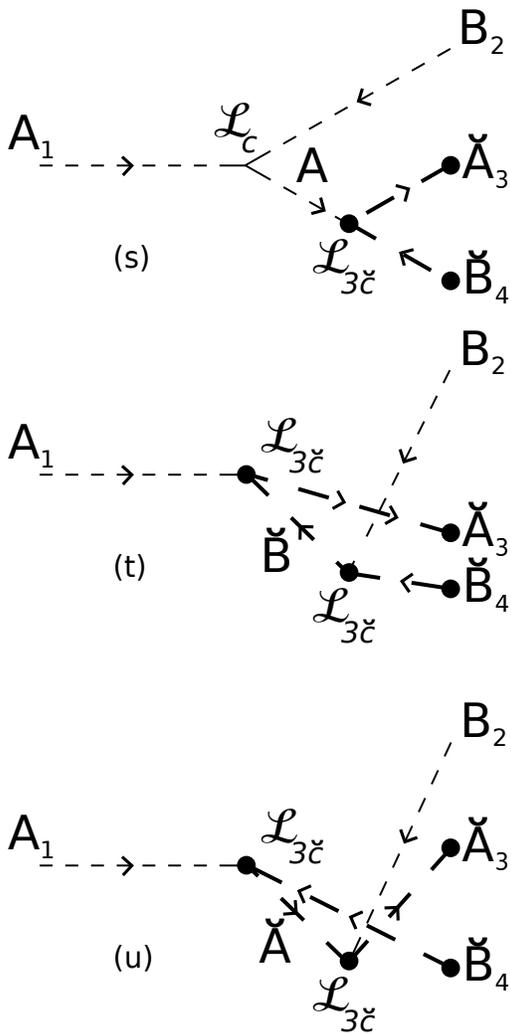}
\caption{\label{fig:epsart} The 3 diagrams for the cascade $A_1 \rightarrow B_2 \breve{A}_3 \breve{B}_4 $ by emission of a particle-antiparticle pair of scalar paraparticles $\breve{A}_3 \breve{B}_4$.}
\end{figure}


\section{Comparison of Predictions for 3 Cases}

To compare the partial widths in the 3 cases, in the Lagrangian densities we assume  the corresponding coupling constants involved in the cascade are equal in the para case and in the two $p=1$ cases of a non-degenerate or a 2-fold degenerate pair. 

For the assumed portal Lagrangian, the scalar pair emission mode $A_1 \rightarrow B_2 \breve{B} _3 \breve{B}_4$ is forbidden through  quadratic order in the Lagrangian densities.  Consequently, when viewed inclusively, the 5 scalar pair cascades from the new $A_1$ boson separate into 3 processes with $A_1 \rightarrow A_2 + \breve{X}$ versus 2 processes with 
$A_1 \rightarrow B_2 + \breve{X}$ because there is the $A_1 \rightarrow A_2 \breve{B}_3 \breve{B}_4 $ cascade. 

The diagrams for these cascade processes with emission of a pair of scalar paraparticles, or of a pair of scalar Bose particles, do have common values for all their respective $c_p$ and $c_d$ statistical factors.  This enables factorization of these common-valued $c_p$ and $c_d$ into overall coefficients.  
When such a factorization occurs, the partial decay widths for the para case versus that for emission of a non-degenerate pair are related by
\ber
d {\Gamma}_{p=2} =   2 ~ | {c_p}/{c_d} |^2 ~  d {\Gamma}_{p=1}
 \eer
where the 2 permutation group basis final states have been summed in the para case.  
From this expression,  two times the partial width is predicted for all paraboson pair cascades versus the Bose case of emission of a non-degenerate pair obeying normal statistics. 

Similarly, the partial width for the para case can be compared with that for the case of emission of a 2-fold degenerate pair  
\ber
d {\Gamma}_{p=2} =    | {c_p}/{c_d} |^2 ~  d {\Gamma}_{deg. pair}
 \eer
The same partial width is predicted for all paraboson pair cascades versus emission of a 2-fold degenerate scalar boson pair. 

In the supersymmetric limit, the mass of the para-Majorana neutrino $\breve{\nu}$ would be the same as that for the scalar paraparticle $ \breve{A}$ and its anti-paraparticle $\breve{B}$,  but in nature the paraparticle spin-$1/2$ and spin-zero masses might be different.  This might enable kinematic separation of a cascade process with 
emission of a pair of para-Majorana neutrinos from one with emission of a pair of scalar paraparticles.     In the case of mass degeneracy of the ${\breve{\nu}} $ and $ {\breve{\mathcal A}} $ particles, generalizations of some of the techniques which exploit the neutrino spin in  $\tau^{-} \rightarrow {\mu}^{-} \nu_{\mu} \nu_{\tau}$ might possibly be used to separate the ${\breve{\nu}}_{\alpha}{\breve{\nu}}_{\beta} $ cascades from the $ {\breve{\mathcal A}} _3 {\breve{\mathcal A}}_4$ cascade.

For  the two para-Majorana neutrino cascades, $A_1 \rightarrow A_2 \breve{\nu}_{\alpha}  \breve{\nu}_{\beta}$ and $A_1 \rightarrow B_2 \breve{\nu}_{\alpha}  \breve{\nu}_{\beta}$,  an overall factorization of  $c_p$ and $c_d$  is also possible,  and  the partial widths in the the para case are twice (the same as) the corresponding partial width for emission of a pair of Fermi Majorana neutrinos (2-fold degenerate Fermi Majorana neutrinos).

\section{Concluding Remarks}

This paper is focused on showing that diagrams, and diagrammatic thinking, can be used in perturbatively analyzing $p=2$ paraparticle processes for an assumed supersymmetric-like ``statistics portal" Lagrangian.  If there is a portal to such paraparticles at the LHC, they might be cascade emitted as a pair of para-Majorana neutrinos as in $A_1 \rightarrow A_2 \breve{\nu}_{\alpha}  \breve{\nu}_{\beta}$ or as a pair of neutral spin-zero paraparticles such as in $A_1 \rightarrow A_2 \breve{A}  \breve{B}$.  The associated connected tree diagrams and their parastatistical factors are obtained above for these 7 cascade processes, through quadratic order in the Lagrangian densities.  For each diagram, these explicit calculations show that  the statistical factor $c_p$ for order $p=2$ parastatistics and the corresponding factor $c_d$ for a non-degenerate or  2-fold degenerate pair which obeys normal statistics, satisfy the easy to remember $c_p=c_d =1$ relation.   

These results complement general quantum field theory results for arbitrary order $p$, including the generalization of the spin-statistics theorem to ``particles of half-integer spin obey parafermi statistics, while particles of integer spin obey parabose statistics" [15].

Certainly the systematic diagrammatic procedure used in this paper, which builds on the successful normal ordering procedure for $p=1$ fields, needs to be shown to 
generalize, especially to higher order non-tree diagram processes involving both $p=1,2$ fields.  However, from the herein calculations, it is noteworthy that the commutator ordering of two parafermi fields in the Lagrangian terms (as dictated by paralocality for observables) is in agreement with the nontrivial respective absence (presence) of coupling in the permutation group basis amplitudes for the    final state  $ | A_l  \breve{\nu}_{\alpha}  \breve{\nu}_{\beta} >_{sym,asym}$ in    $A_1 \rightarrow A_2 \breve{\nu}_{\alpha}  \breve{\nu}_{\beta}$ and $A_1 \rightarrow B_2 \breve{\nu}_{\alpha}  \breve{\nu}_{\beta}$. This occurs diagram by diagram.  Likewise, the anticommutator ordering of two parabose fields
in the Lagrangian terms is also in agreement with the absence (presence) of coupling, again diagram by diagram, in the permutation group basis amplitudes for two final scalar parabosons in a totally antisymmetic (symmetric) final state in the 5 cascade processes,   $A_1 \rightarrow A_2 \breve{A}  \breve{B}$, $A_1 \rightarrow B_2 \breve{A}_3 \breve{A}_4$,  $\cdots$.

While the permutation group basis is always physically required in constructing the associated physical amplitudes for all parabosons or all parafermions in the external final (initial) states, the convenient usage of the occupation number basis in the calculations in this paper also generalizes to more than two final paraparticles:  

In the case of more than two parabosons, the central idea of only evaluating one occupation number basis amplitude for each diagram works.  For instance, for 4 final parabosons of order $p=2$, the totally symmetric final state which uses the totally symmetric bracket 
$\{a^{\dag}  b^{\dag} c^{\dag} d^{\dag} \} _+$ is
\ber
 \frac {1}{4  \sqrt{6} } \{a^{\dag}  b^{\dag} c^{\dag} d^{\dag} \} _+ |0> =  \frac {1}{  \sqrt{6} } ( a^{\dag}  b^{\dag} c^{\dag} d^{\dag} + b^{\dag}  a^{\dag} d^{\dag} c^{\dag}+ \nonumber \\
 a^{\dag}  d^{\dag} b^{\dag} c^{\dag} + 
 d^{\dag}  a^{\dag} c^{\dag} b^{\dag}+ a^{\dag}  c^{\dag} d^{\dag} b^{\dag} + b^{\dag}  d^{\dag} c^{\dag} a^{\dag}) |0>
 \eer 
The state has 6 independent orthogonal terms.  However, for each diagram only one amplitude needs to be calculated in the occupation number basis, for instance, the amplitude for the $a^{\dag}  b^{\dag} c^{\dag} d^{\dag}  |0>$ term.   The other amplitudes easily follow by permutations of the mode labels.  

In (23) and in the other state expressions in this section, the states are normalized, but with the $\frac{1}{\sqrt{2}}$ factor for each paraparticle omitted. Note, the reordering relations [16] of Appendix E  must first be used to reduce the $4! = 24$ terms from the totally symmetric left-hand-side of (23), to 6 independent terms.   There are 6 independent orthogonal terms because the sum of the dimensions of the three permutation group irreducible representations for 4 parabosons is 6.  This totally symmetric permutation group row representation is 1-dimensional and has an eigenvalue of $6$ for $  {\overline P} _{sum} $. This directly physical $  {\overline P} _{sum} $ operator is the sum of the pair particle-exchange operators, see (A6) for three parabosons in  Appendix A.   

This single occupation number basis amplitude for the $a^{\dag}  b^{\dag} c^{\dag} d^{\dag}  |0>$ term then also suffices for construction of the permutation basis amplitude for each of the other two permutation irreducibles.  For 
the L-shaped representation with dimension  $3$ and an eigenvalue of $2$ for $  {\overline P} _{sum}$, there is the eigenvector $\frac{1}{\sqrt{2}} ({a}^{\dag} {b}^{\dag} {c}^{\dag} d^{\dag}- {b}^{\dag} {a}^{\dag} {d}^{\dag} c^{\dag}) |0> $.  Finally, for  
the box-shaped representation with dimension  $2$ and an eigenvalue of $0$ for $  {\overline P} _{sum} $, there is the similar eigenvector $\frac{1}{2} ([   {a}^{\dag} {b}^{\dag} {c}^{\dag}d^{\dag} + {b}^{\dag} {a}^{\dag} {d}^{\dag} {c}^{\dag} ] - 
   [   {a}^{\dag} {d}^{\dag} {b}^{\dag} {c}^{\dag}+ {d}^{\dag} {a}^{\dag} {c}^{\dag} {b}^{\dag}         ] ) |0>$.  
   
  In summary, for 4 final parabosons for each diagram there is one occupation number basis amplitude which requires evaluation.  By permuting the external particle mode labels, this single amplitude then gives by superposition the three amplitudes which each correspond to the three distinct permutation group basis irreducible representations (the totally symmetric, the L-shaped, and the box-shaped).

For 4 final parafermions the amplitude evaluation procedure and the state decompositions are very similar with the permutation group basis irreducibles having the same dimension but opposite sign of $  {\overline P} _{sum} $ eigenvalues versus 4 parabosons.  Again, only one occupation basis amplitude needs to be evaluated for each diagram.
Appendix F contains independent basis states and $  {\overline P} _{sum} $ eigenvalues for up to 4 parabosons (parafermions).


\appendix
\section{Tri-Linear Relations for a ``$p=2$ Family" of Parafields }

In the calculations of the cascade matrix elements, the following tri-linear relations [1] for a ``$p=2$ family" of parafields, $ \breve{\mathcal A}(y)$  and  $ \breve{\xi}(x) $,  are used with parabose operators denoted with Roman letters and parafermi operators denoted with Greek letters.  In the supersymmetric-like model in the present paper, there are of course an equal number of parabose and parafermi degrees of freedom.  The parastatistics term ``$p=2$ family"  means that all the fields in the family mutually obey these tri-linear relations [7].  The fields,  $ \breve{\xi}(x) $ and  $ \breve{\mathcal A}(y)$, have their usual covariant momentum-expansions and normalizations in terms of these creation and annihilation operators, see (C1) and (C2) below.   In the arbitrary $p$ order tri-linear relations, versus the following $p=2$ tri-linear relations, there are twice as many terms on the left-hand side of each relation due to an additional overall commutator ordering [3].

The mode index $k,l,m$ includes the momentum components, and the helicity components for the para-Majorana field $\breve{\xi}$, and  the $\breve{A}$, $\breve{B}$ particle-antiparticle distinction for the $\breve{\mathcal{A}}$ complex field.   For instance, in the tri-linear relations below for the para-Majorana operators, the generalized Kronecker delta is $ \delta_{lm}=\delta_{\lambda_l \lambda_m}  \delta^{(3)} (\vec {p}_l - \vec{p}_m)$.  Here, for clarity, we omit/suppress a possible but awkward ``breve" accent which might be put on top of each of the creation and annihilation operators.  

Several simple patterns are apparent:  As for the usual $p=1$ bi-linear relations, in each relation the left-hand-side has the second term with the three operators written in opposite cyclic-order to that of the first term.  The second term has a plus (minus) sign when mostly parabosons (parafermions) occur in the tri-linear relation.  On the right-hand-side, the existence of a Kronecker delta term, and its sign, corresponds to an $a_k a^{\dag}_l$ or $ {\alpha}_k {\alpha}^{\dag}_l$ adjacent-pair factor from the left-hand-side.  The tri-linear relations maintain the associated odd (even) ``place positions" [14] of both the mode and also of the parafermi/parabose labeling of the operators, whether reading left-to-right, or right-to-left.  These simple properties also occur in the adjointed relations.  The normalization of these $p=2$ relations corresponds to that of the tri-linear relations for arbitrary $p$ parastatistics [1, 3].  The usual $p=1$ creation and annihilation operators for boson fields, such as the scalar complex field $\mathcal{A}$,  commute  with these $p=2$ operators and those for fermion fields, such as the Majorana spin-$1/2$ field $\xi$, commute (anticommute) with the parabosons (parafermions).

For all parabosons (Roman letters):  
\ber
a_k a_l a_m - a_m a_l a_k =0,   \nonumber \\
a_k a_l a_m^{\dag} - a_m^{\dag} a_l a_k = 2 \delta_{lm} a_k \nonumber \\
a_k a_l^{\dag} a_m - a_m a_l^{\dag} a_k = 2 \delta_{kl} a_m -2 \delta_{ml} a_k 
\eer
For all parafermions (Greek letters):
\ber
{\alpha}_k {\alpha}_l {\alpha}_m + {\alpha}_m {\alpha}_l {\alpha}_k =0,   \nonumber \\
{\alpha}_k {\alpha}_l {\alpha}_m^{\dag} + {\alpha}_m^{\dag} {\alpha}_l {\alpha}_k = 2 \delta_{lm} {\alpha}_k \nonumber \\
{\alpha}_k {\alpha}_l^{\dag} {\alpha}_m + {\alpha}_m {\alpha}_l^{\dag} {\alpha}_k = 2 \delta_{kl}{\alpha}_m +2 \delta_{ml} {\alpha}_k 
\eer
For two parabosons and one parafermion:
\ber
a_k a_l {\beta}_m - {\beta}_m a_l a_k =0,   \nonumber \\
a_k {\beta}_l a_m - a_m {\beta}_l a_k =0,   \nonumber \\
a_k a_l {\beta}^{\dag}_m - {\beta}^{\dag}_m a_l a_k =0,   \nonumber \\
a_k {\beta}_l a^{\dag}_m - a^{\dag}_m {\beta}_l a_k =0,   \nonumber \\
{\beta}_k a_l a_m^{\dag} - a_m^{\dag} a_l {\beta}_k = 2 \delta_{lm} {\beta}_k \nonumber \\
a_k a_l^{\dag} {\beta}_m - {\beta}_m a_l^{\dag} a_k = 2 \delta_{kl} {\beta}_m \nonumber \\
a_k {\beta}_l^{\dag} a_m - a_m {\beta}_l^{\dag} a_k = 0 
\eer

For two parafermions and one paraboson:
\ber{\alpha}_k {\alpha}_l b_m + b_m {\alpha}_l {\alpha}_k =0,   \nonumber \\
{\alpha}_k b_l {\alpha}_m + {\alpha}_m b_l {\alpha}_k =0,   \nonumber \\
{\alpha}_k {\alpha}_l b^{\dag}_m + b^{\dag}_m {\alpha}_l {\alpha}_k =0,   \nonumber \\
{\alpha}_k b_l {\alpha}^{\dag}_m + {\alpha}^{\dag}_m b_l {\alpha}_k =0,   \nonumber \\
b_k {\alpha}_l {\alpha}^{\dag}_m+ {\alpha}^{\dag}_m {\alpha}_l b_k = 2 \delta_{lm} b_k \nonumber \\
{\alpha}_k {\alpha}^{\dag}_l b_m + b_m {\alpha}^{\dag}_l {\alpha}_k = 2 \delta_{kl} {b}_m \nonumber \\
{\alpha}_k b_l^{\dag} {\alpha}_m + {\alpha}_m b_l^{\dag} {\alpha}_k = 0 
\eer

In this arbitrary $p$ order normalization, the important associated vacuum conditions for any mode indices $k, l $ are
\ber
a_k | 0 > = \alpha_l |0> =0 , \; <0|0> =1    \nonumber \\
a_k a_l^{\dag}  | 0 > =   {\alpha}_k {\alpha}_l^{\dag}                      | 0 > = 2 \delta_{kl}  | 0 >  
\eer
 and $ a_k  {\alpha}_l^{\dag} | 0 > =   {\alpha}_k       a_l^{\dag}                | 0 > =0$.  
 The parabose and parafermi number mode operators are respectively
  $ N_k= \frac{1}{2} \{ a_k^{\dag}, a_k \}  - 1   $  and  $
\mathcal{ N}_k= \frac{1}{2} [ \alpha_k^{\dag}, {\alpha}_k ] +  1 $.  

For $p$ order parastatistics, in the vacuum conditions (A5), there is the substitution $2 \rightarrow p $, and in the number operators the ending terms $ \mp 1 \rightarrow  \mp \frac{p}{2} $, so the important zero point energies scale with the order $p$.
Associated with (A5), for $p=2$ there is an extra $(\frac{1}{\sqrt{2}})$ factor for each paraparticle in an external state.  Thereby, the scattering matrix, and associated in-going and out-going particle fluxes,  have a common ``particle density per unit volume" normalizaton [11] for all external particles whether of order $p=2$ or $p=1$.

For an initial state, final state, or observable expressed as a function of creation and annihilation operators, the directly physical ``particle permutations" 
are products of the pair particle-exchange operators $ \overline{P}_{i,j} = \overline{P}_{j,i}$ which exchange the $i$ and $j$ identical particles, so $a_i \leftrightarrow a_j$ or 
$a_i^{\dag} \leftrightarrow a_j^{\dag}$.  As in [14], in the present paper these operators are denoted with an ``overbar."  Instead, ``place permutations" are products of the pair place-exchange operators $P_{r,s}= P_{s,r}$ which exchange the occupants of positions $r$ and $s$ in a creation and annihilation operator expression regardless of the identity of the  occupants.

In $p=2$ parastatistics, unlike for $p=1$ quanta, the external ``permutation group basis" states are in general not eigenstates of the pair particle-exchange operators $ \overline{P}_{i,j} $.
Indeed, for two identical paraparticles, the external states are pair particle-exchange eigenstates as in (19) for parafermions and in (20) for parabosons.  For three identical parabosons, the totally symmetric 1-dimensional external state is also even under each of the three particle-exchanges $ \overline{P}_{i,j} $.  However, the three identical paraboson state corresponding to the 2-dimensional  L-shaped representation has basis vectors which are not eigenstates of the three   $ \overline{P}_{i,j} $.  For this mixed representation and using the reordering relations of Appendix E, this is apparent because the two independent basis vectors can be chosen as  $\frac{1}{\sqrt{2}} ({a}^{\dag} {b}^{\dag} {c}^{\dag} - {b}^{\dag} {c}^{\dag} {a}^{\dag} ) |0> $  and $ \frac{1}{\sqrt{6}} ({a}^{\dag} {b}^{\dag} {c}^{\dag} + {b}^{\dag} {c}^{\dag} {a}^{\dag} - 2 {c}^{\dag} {a}^{\dag} {b}^{\dag} ) |0> $. 
 When one of the three $ \overline{P}_{i,j}$ acts on either one of these two basis vectors, it gives a linear combination of them.  More generally, acting with any of the $ \overline{P}_{i,j}$ on a state in an $n$ particle irreducible representation of the permutation group preserves its irreducible representation.

The sum of the three particle-exchanges $ \overline{P}_{i,j}$ which we denote   $  {\overline P} _{sum} $ 
\ber
 {\overline P} _{sum} = \overline{P}_{a,b}+\overline{P}_{b,c}+\overline{P}_{c,a}
\eer
has respective eigenvalues $3$ and $0$ for these two parabose representations,  (row) and (L-shape), and so it can also be used to label them.  Similarly, we find that states in $n$ dimensional parabose representations are eigenstates of $  {\overline P} _{sum} $ (at least thru the four 6 paraboson irreducibles).  For the $n$-dimensional totally symmetric representations, the eigenvalue $n(n-1)/2$ is equal to the number of pair particle-exchange operators.  
For states of identical parafermions, the eigenvalues of $  {\overline P} _{sum} $ for the corresponding irreducible representations are negative.  A diagonal mirror reflection of rows and columns transforms a paraboson irreducible representation to a corresponding parafermi irreducible representation.  While the dimension of the permutation group representation provides one label for the irreducible representation for $p=2 $ external state, the 2-particle state shows that the $  {\overline P} _{sum} $ eigenvalue is also required for a unique labeling.  Also for the 6 paraboson state, the $  {\overline P} _{sum} $ eigenvalues of 9 (L-shape) and 3 (box-shape) of  distinguish these representations which are both 5-dimensional.

\section{Paralocality, Green Components, and Possible Additional Interaction Terms }

At the beginning of Section 2, in the construction of the supersymmetric-like portal Lagrangian densities, ``paralocality" is used.  It is a generalization of locality for parafields [3]. The allowed forms of paraparticle couplings arise as a consequence of the tri-linear commutation relations and the assumed locality condition.  By locality, for the two obserables $ {\mathcal O}(x)$ and  $ {\mathcal O}^ {'} (y)$, their commutator must vanish
\ber
[ {\mathcal O}(x),{\mathcal O}^ {'} (y) ] =0 ,  \;   x \sim y
\eer
when points $x$ and $y$ are spacelike separated (denoted by the symbol $\sim$).  In the interaction picture, $ {\mathcal O}(x)$ and  $ {\mathcal O}^ {'} (y)$ are polynomial functions of the free parafield operators $ \breve { \phi} _i (x)$ which act on the vacuum of the physical Hilbert space. In arbitrary order $p$ parastatistics, ``paralocality" holds when (B1) is valid in the larger Hilbert space of the Green component fields $  {\breve { \phi} ^{(a)}_i (x)}$ defined by the expansion
\ber
\breve { \phi} _i (x) = { \sum_{a=1}^p } \;  { \breve { \phi}^{(a)}_i (x)}.
\eer
where $a$ is the Green index.
For parabosons (parafermions) these Green component fields with the same Green index,  $  {\breve { \phi} ^{(a)}_i (x)}$ and
 $   {\breve { \phi}_i } ^{(a) {\dag} }(y) $, 
 obey the usual Bose (Fermi) commutation relations, but  anticommute (commute) with all $  {\breve { \phi} ^{(b)}_i (z)}$  and $ {\breve { \phi}_i } ^{(b) {\dag} }(z)  $ for $a \neq b$.  In an ``order $p$ family" of parafields, a parabose Green component and a parafermi Green component have the same commutation pattern as two paraboson Green components.  Green components were introduced in [1].   In [12, 7], it is shown that locality implies paralocality. 
 
While in the perturbative calculations in this paper for order $p=2$ we do not expand in Green components, they are very convenient tools for analysis and for checking.    Historically, Green components have been exceptionally useful in developing and understanding fields and quanta obeying parastatistics, especially for arbitrary $p$ order.   Their underlying presence in parastatistics is a strong physics/mathematics motivation for the consideration, throughout the present paper, of the comparison with cascade emission of a 2-fold degenerate pair of particles which obey normal statistics.
 
In above portal Lagrangian densities, we do not consider possible additional ``second unit observables" which are allowed by paralocality.  A generic example, in terms of paraparticle creation or annihilation operators denoted by a $ \hat{c}_i$ (the hat accent denotes $ {c}_i$ or  $ {c}^{\dag}_i$ ) is 
 \ber
 \vdots { [ \hat{c}_1, \hat{c}_ 2, \cdots  \hat{c}_n ]_{\mp}}  \vdots \equiv n! {\sum_ { a_1,a_2, \cdots, a_n} ^{} }   {\hat{c}_1}^{(a_1)}  {\hat{c}_ 2}^{(a_2)} \cdots  {\hat{c}_n}^{(a_n)}
\nonumber \\
\eer
where the summation is over all different values of the Green indices $a_1,a_2, \cdots, a_n $.  
As denoted by its redundant $\mp$ subscript, this ``dotted bracket" or ``second unit observable" is totally antisymmetric  (symmetric) with respect to the labels $1,2, \dots n$  in the case of all parabose (parafermi) operators and it respectively vanishes for $n > p$, which is another meaning for the order $p$.    In permutations for (B3), it is understood that the dagger, or no-dagger, on $ \hat{c}_i$, moves with the subscript $i$.

In the context of deriving the most general selection rules for particles obeying parastatistics, these second unit observables were introduced in [3],  (see earlier [17]).    Such additional terms are treated in detail in Ref. [7].  However, these additional terms are forbidden if either there is the stronger locality condition 
 \ber
[\breve { \phi} _i (x) ,{\mathcal O}^ {'} (y) ] =0 ,  \;   x \sim y
\eer
or if there is a global symmetry such that the Green indices transform under O(2) or U(2), instead of the smaller SO(2) or SU(2).   These two locality conditions, (B1) and (B4), are equivalent for ordinary fermions and bosons but are not for even-valued orders of $p$, see [7].

\section{Evaluation of  $ p=2 $ Matrix Elements }  
 Some care is needed in the evaluation of $p=2$ matrix elements because of the factor of $2$ in the vacuum condition $c_k c_l^{\dag}  | 0 > = 2 \delta_{kl}  | 0 >  $.  In the ``arbitrary $p$ normalization" of [3] which is used in this paper, a factor of  $\frac {1} {\sqrt{2}}$ does not occur in a parafield's momentum-expansion such as for the complex spin-zero parabose $ \breve{\mathcal A}$ field
\ber
    \breve{\mathcal A} (x) = \int \frac{d^3 q} { \sqrt{  (2 \pi)^{3}   {2 \omega_q} }   } ( \breve{ A}(\vec{q}) e^{-i q \cdot x} + \breve{B}^{\dag}(\vec{q}) e^{i q \cdot x} ) \nonumber \\
\eer
Consequently, $  \breve{\mathcal A} (x)  \breve{\mathcal A}^{\dag} (y)|0> = 2 \delta^{(3)} ( \vec{x} - \vec{y})|0> $,  and $ 2  i {\Delta}_F (q) = 2  {i} /  (q^2 - {m}^2 + i \epsilon )$ in momentum space corresponds to 
$ <0 | T ( \breve{\mathcal A} (x^\prime)  \breve{\mathcal A}^{\dag} (x)   ) |0> $, with these three $2$'s  replaced by $p$'s for $p$ order parastatistics.   Similarly, there are corresponding factors of $2$ occurring for the  Majorana spin-$1/2$ parafermi  $\breve{\xi}$ field  
\ber
\breve{\xi}_A (x) = \sum_{\lambda = \pm 1/2}  \int \frac{d^3 p} { \sqrt{  (2 \pi)^{3}   {2 \omega_p} } } ( x_A (\vec{p}, \lambda)  \breve{\alpha}_{\lambda}(\vec{p})e^{-i p \cdot x}  \nonumber \\
+ y_A (\vec{p}, \lambda)  \breve{\alpha}_{\lambda}^{\dag}(\vec{p}) e^{i p \cdot x} )  \nonumber \\
\eer
with $ \bar{\breve{\xi}}_{\dot{A}} (x) = {\breve{\xi}}^{\dag }_{\dot{A}} (x) = ({\breve{\xi}}_{{A}} (x) )^{\dag}$.  For essential properties of the commuting two-component Weyl spinor wave functions $ x_A$ and $y_A $  see DHM [10].

\section{Same Parastatistical Factors for the $p=2$ Normalization of Green and Volkov} 

A simple, but very partial, working check of the above perturbative evaluations is to use the alternate
 $p=2$ normalization of Green and Volkov [1] for the tri-linear relations of Appendix A.  If it is used, the same parastatistical factor $c_p$ is obtained for each diagram for these 7 cascade processes.  The differences are:  (i) For both the $p=2$ parabosons and parafermions, the quanta operators $ c_i \rightarrow  {\sqrt{2}} d_i$ in the tri-linear relations and in the vacuum conditions (A5) of Appendix A.  (ii) Each of the portal Lagrangian densities is quadratic in the parafields, so each of the coupling constants in (4-9) must be rescaled by $g_i \rightarrow 2 g_i$.  The momentum-expansions in Appendix C are still used without additional factors of ${\sqrt{2}}$, so there is no change in the normalization of the parafields versus the arbitrary $p$ normalization used in this paper.  Conversely, if this Green-Volkov $p=2$ normalization is used, then the arbitrary $p$ normalization is obtained by the substitution $ d_i \rightarrow \frac {1} {\sqrt{2}} c_i$ but with the extra $ \frac {1} {\sqrt{2}} $ factor, which would occur in each parafield momentum-expansion, instead moved out to be with the coupling constant in the Lagrangian density.

\section{Reordering Relations for $p=2$}  

With upper (lower) signs for parabose (parafermi) creation operators, 
the reordering relations [16] are 
\ber
{a}^{\dag} {b}^{\dag}  {c}^{\dag}   = \pm   {c}^{\dag}  {b}^{\dag} {a}^{\dag}    \nonumber \\
( {a}^{\dag} {b}^{\dag} )  ( {c}^{\dag}  {d}^{\dag} ) =   ( {c}^{\dag}  {d}^{\dag} ) ( {a}^{\dag} {b}^{\dag} ) =   \nonumber \\
\pm ( {c}^{\dag}{b}^{\dag} )  ( {a}^{\dag}  {d}^{\dag} ) 
=  \pm ( {a}^{\dag}  {d}^{\dag} )  ( {c}^{\dag} {b}^{\dag} )  
\eer
The 3 operator relation is a cyclic one.  In the 4 operator relations, the unnecessary pairing parentheses are for displaying the pairing patterns.  These patterns are that in reordering: (i) the even (odd)  place positions are maintained, whether reading from the left or right, and (ii) each single exchange of $a^{\dag} \leftrightarrow c^{\dag}$ or $b^{\dag} \leftrightarrow d^{\dag}$ gives a $\pm$ sign.   Both patterns are also in the 3 operator relation.
The annihilation operators satisfy the same relations (remove daggers). 

As discussed in the last several paragraphs of the text, these relations are particularly useful in the construction of the independent orthogonal terms needed in the external states in the permutation group basis.    

These reordering relations follow from the first lines of (A1) and (A2).  For 3 mixed parabose and parafermi, all creation (annihilation) operators there are also reordering relations corresponding to (E1) which follow from the first two lines of (A3) and of (A4):  These analogous 3 operator relations to (E1) are also cyclic and hold with the $\pm$ sign for mostly parabosons (parafermions).  The appropriate signs in the 4 or more operator relations for any mixture of parabosons and parafermions then follow iteratively. 

\section{Independent Basis States and $  {\overline P} _{sum} $ Eigenvalues for Up to 4 Parabosons (Parafermions) }  

In this appendix ``eigenvalue" for an $n$ parabose or parafermi state means the eigenvalue of the $  {\overline P} _{sum} $ operator which is the sum of the pair particle-exchange $ \overline{P}_{i,j}$ operators, see end of Appendix A.  As in the concluding Sec.IV, in this appendix we are suppressing the extra $\frac{1}{\sqrt{2}}$ normalization factor for each paraparticle.  When these extra factors are included, each state is properly normalized. Each term in these states is independent and orthogonal because the reordering relations of Appendix E have already been used.

For $2$ parabosons, both of the two irreducible representations are 1-dimensional. These correspond to the totally symmetric row (antisymmetric column) in a Young diagram, and have basis states
\ber
\frac{1}{\sqrt{2}} [{a}^{\dag} ,{b}^{\dag} ]_{\pm} |0>  
\eer
with  eigenvalues $\pm 1$.  The bracket's $\pm$ subscript denotes anticommutator (commutator). 

For $3$ parabosons, there is the totally symmetric $1$-dimensional row representation with
\ber
\frac{1}{\sqrt{3}} ({a}^{\dag} {b}^{\dag} {c}^{\dag} + {b}^{\dag} {c}^{\dag} {a}^{\dag} +{c}^{\dag} {a}^{\dag} {b}^{\dag} ) |0>  
\eer
and eigenvalue $3$. For the 2-dimensional, L-shaped permutation group representation with an eigenvalue equal to $0$, two basis vectors are
\ber
\frac{1}{\sqrt{2}} ({a}^{\dag} {b}^{\dag} {c}^{\dag} - {b}^{\dag} {c}^{\dag} {a}^{\dag} ) |0>  \nonumber \\
\frac{1}{\sqrt{6}} ({a}^{\dag} {b}^{\dag} {c}^{\dag} + {b}^{\dag} {c}^{\dag} {a}^{\dag} - 2 {c}^{\dag} {a}^{\dag} {b}^{\dag} ) |0> 
\eer

For $4$ parabosons, there is the totally symmetric 1-dimensional row representation with
\ber
\frac {1}{  \sqrt{6} } ( a^{\dag}  b^{\dag} c^{\dag} d^{\dag} + b^{\dag}  a^{\dag} d^{\dag} c^{\dag}+ \nonumber \\
 a^{\dag}  d^{\dag} b^{\dag} c^{\dag} + 
 d^{\dag}  a^{\dag} c^{\dag} b^{\dag}+ a^{\dag}  c^{\dag} d^{\dag} b^{\dag} + b^{\dag}  d^{\dag} c^{\dag} a^{\dag}) |0>
 \eer 
and eigenvalue $6$.  For the 3-dimensional, L-shaped representation with an eigenvalue equal to $2$, three basis vectors are
\ber
\frac{1}{\sqrt{2}} ({a}^{\dag} {b}^{\dag} {c}^{\dag} d^{\dag}- {b}^{\dag} {a}^{\dag} {d}^{\dag} c^{\dag}) |0>  \nonumber \\
\frac{1}{\sqrt{2}} (  {a}^{\dag} {d}^{\dag} {b}^{\dag} c^{\dag} - {d}^{\dag} {a}^{\dag} {c}^{\dag} b^{\dag} ) |0>  \nonumber \\
\frac{1}{\sqrt{2}} (  {a}^{\dag} {c}^{\dag} {d}^{\dag} b^{\dag} - {b}^{\dag} {d}^{\dag} {c}^{\dag} a^{\dag}) |0>  
\eer
For the 2-dimensional, box-shaped representation with an eigenvalue equal to $0$, two basis vectors are
\ber
\frac{1}{2} ([   {a}^{\dag} {b}^{\dag} {c}^{\dag}d^{\dag} + {b}^{\dag} {a}^{\dag} {d}^{\dag} {c}^{\dag} ] - 
   [   {a}^{\dag} {d}^{\dag} {b}^{\dag} {c}^{\dag}+ {d}^{\dag} {a}^{\dag} {c}^{\dag} {b}^{\dag}         ] ) |0>  \nonumber \\
\frac{1}{ 2 \sqrt{3} }  ( [   {a}^{\dag} {b}^{\dag} {c}^{\dag} {d}^{\dag} + {b}^{\dag} {a}^{\dag} {d}^{\dag} {c}^{\dag} ] + 
   [ {a}^{\dag} {d}^{\dag} {b}^{\dag} {c}^{\dag}+ {d}^{\dag} {a}^{\dag} {c}^{\dag} {b}^{\dag} ]   +  \nonumber \\
   -  2  [  {a}^{\dag} {c}^{\dag} {d}^{\dag} {b}^{\dag} + {b}^{\dag} {d}^{\dag} {c}^{\dag} {a}^{\dag} ]      ) |0>   \; \; \; 
  \eer
  
For the $n$ parafermion states, first recall that a diagonal mirror reflection of rows and columns transforms a paraboson Young diagram irreducible representation to the corresponding parafermi representation.   The $  {\overline P} _{sum} $ eigenvalue is minus that for the corresponding paraboson representation.    For the 2 parafermion irreducible representations, there are the (F1) basis states but with Greek letters  per the notation in Appendix A.  

For 3 and 4 parafermions, the basis vectors are the same as above but with a change from Roman to Greek letters.   In the above basis vectors the expected different signs for parafermions, versus parabosons, have already been absorbed by using the reordering relations of Appendix E to reduce the basis vector expressions to the displayed independent orthogonal terms.  For instance, if for 3 parafermions, one constructs the 1-dimensional totally antisymmetric representation by first explicitly writing out the $3!$ terms, the re-ordering relations can be used to reduce it to (F2), but with Greek letters. 



\end{document}